\documentclass[aps,prl,twocolumn]{revtex4}

\usepackage{graphicx}

\begin{document}

\title{Reaction Pathways Based on the Gradient of the Mean First-Passage Time}

\author{Sanghyun Park}

\author{Klaus Schulten}

\affiliation{Beckman Institute and Department of Physics, 
University of Illinois at Urbana-Champaign, Urbana, Illinois 61801}

\begin{abstract}
Finding representative reaction pathways is necessary for 
understanding mechanisms of molecular processes, but is considered to be
extremely challenging.
We propose a new method to construct reaction paths 
based on mean first-passage times.
This approach incorporates information of all possible reaction events 
as well as the effect of temperature.
The method is applied to exemplary reactions in a continuous and in a 
discrete setting.
The suggested approach holds great promise for large reaction networks
that are completely characterized by the method through a pathway graph.
\end{abstract}

\maketitle

\paragraph{\textbf{Introduction.}}

In many physical, chemical, or biological reactions,
initial (reactant) and final (product) states are known 
but reaction pathways are not.
Examples range in complexity from single-particle Brownian motion to 
conformational changes of proteins, such as protein 
folding~\cite{Pande1998,Eastman2001}.
Finding reaction pathways is one of the most fundamental challenges 
in chemistry and molecular biology\cite{Elber1996,Straub2001}.
It is important for understanding reaction mechanisms 
and for the calculation of reaction rates~\cite{Hanggi1990}.

In most cases of interest, reactions take place at finite
temperature and therefore are stochastic: 
every reaction event follows a different path
and takes a different amount of time.
Among all possible paths from reactant to product,
one seeks the path that best characterizes the reaction~\cite{note:sampling}.
The selection criteria are twofold. 
First, one seeks \textit{minimal} paths, free of unnecessary 
fluctuations. 
Second, and more important, one seeks paths that are \textit{representative}
of all reaction events so that individual reaction events can be considered 
to be following noisy paths around them.
It is, however, challenging to formulate these criteria rigorously.

The most widely used formulation of reaction path is probably
the steepest-descent path, 
which is constructed by identifying saddle points of the potential energy
surface and then following the steepest descent from the saddle points
such that energy barriers along the path are minimized~\cite{Elber1996}.
The steepest-descent path, however, does not involve temperature.
Since reactions are driven by thermal fluctuations, 
reaction paths should depend on temperature.
For example, if there is a direct path with high energy
barriers and a roundabout path with low energy barriers,
at a temperature higher than the barriers
reaction will occur most likely along the direct path
rather than the roundabout path while the steepest-descent path will be the
roundabout path regardless of temperature.

Understanding this drawback of the steepest-descent path has led to 
alternative formulations of reaction path.
One approach is to select the path of 
maximum flux~\cite{Berkowitz1983,Huo1997}.
Another formulation focuses on most probable paths~\cite{Pratt1986,Elber2000}.
In this approach, an ensemble of reaction events of a $\textit{fixed}$ time
interval is considered.
A probability is then assigned to each event, and the path followed by 
the most probable event is taken as the reaction path~\cite{note:bvp}.

The above methods succeeded to a certain extent in elucidating
reaction mechanisms, 
but they do not fully satisfy the criterion of representativity.
The method of most probable path comes closest to satisfying the criterion, 
but it is not clear how to choose the time interval beforehand 
and whether an ensemble of reaction events of a single time interval 
suffices to represent the reaction.

In this paper we present a new formulation of reaction path. 
While previous approaches attempted to quantify \textit{paths}, 
we use the concept of reaction coordinate which quantifies \textit{states}.
Reaction coordinate is a function that describes 
where in the progress of reaction a state is located. 
The most natural measure of the progress of reaction is provided by the mean
first-passage time (MFPT), namely the average amount of time 
that the system starting from the state takes to reach the given product.
The MFPT depends on the energy landscape, the temperature, 
as well as the boundary conditions, and most important, it is an average over 
all reaction events.
Surprisingly, to our knowledge, the MFPT has never been used as a reaction
coordinate before.
Once the MFPT is determined, one can choose as reaction paths
the paths along which the MFPT decreases most rapidly, 
which complies with the criterion of minimal path.

\paragraph{\textbf{Theory.}}

\begin{figure}
  \includegraphics{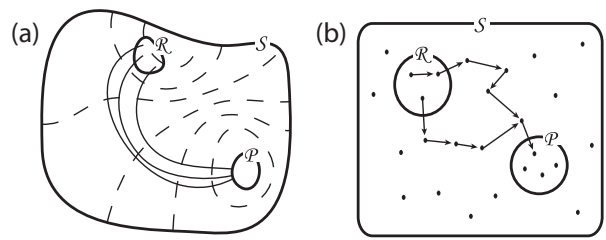}
\caption{\label{fig/picture} Schematic pictures of reaction paths.
$\mathcal{S}$ is the set of all accessible states,
$\mathcal{R}$ the reactant, and $\mathcal{P}$ the product.
(a) A continuous system. Dashed lines are contours of the reaction coordinate
(the MFPT to the product $\mathcal{P}$).
In a Cartesian coordinate system, 
they are orthogonal to the reaction paths.
(b) A discrete system. Dots are the accessible states.}
\end{figure}

A practical scheme of determining reaction coordinates and 
reaction paths based on MFPTs can indeed be stated, as described below and
illustrated in Fig.~\ref{fig/picture}.
Consider a system undergoing a stochastic reaction and let 
$\mathcal{S}$ be the set of all states that the system can access.
If $\mathcal{S}$ is continuous the reaction can be described
by a Fokker-Planck-type equation, and if $\mathcal{S}$ is discrete 
it can be described by a master equation~\cite{Gardiner1985}.
The reactant and the product are specified by disjoint subsets,
$\mathcal{R}$ and $\mathcal{P}$ respectively, of $\mathcal{S}$.
The reaction can be considered a first-passage process~\cite{Redner2001} 
because a reaction event ends as soon as the system reaches 
any state in $\mathcal{P}$.
The MFPT $\tau(\mathbf{x})$ from state $\mathbf{x}$ to the product
$\mathcal{P}$ is then calculated for all states $\mathbf{x}$
in $\mathcal{S}$ 
(this involves solving an inhomogeneous partial differential equation 
when $\mathcal{S}$ is continuous and inverting a transition matrix 
when discrete~\cite{MFPT,Bauer1988}, as demonstrated later)
and is used as a reaction coordinate.
The location of the reactant set $\mathcal{R}$ does not affect the calculation
of the reaction coordinate $\tau(\mathbf{x})$; 
it is involved only in determining reaction paths.

The scheme of constructing reaction paths from MFPTs depends on the character 
of the set $\mathcal{S}$.
When $\mathcal{S}$ is continuous and described by a Cartesian coordinate
system, reaction paths are constructed 
following the direction of $-\nabla\tau$, 
along which $\tau$ decreases most rapidly.
Thus, a reaction path $\mathbf{x}(l)$, 
parameterized through an arc length $l$, satisfies
\begin{equation}
\label{rp_Cartesian}
  {dx_i\over dl} 
  = v_i\Big(\sum\nolimits_j v_j v_j\Big)^{-1/2}\;,\quad
  v_i = -{\partial\tau\over\partial x_i}\;.
\end{equation}
Often, reactions are better described with non-Cartesian
coordinates such as angles.
In such cases reaction paths can be determined via a transformation to
a Cartesian coordinate system, and the resulting equation of reaction path is
\begin{equation}
\label{rp_nonCartesian}
  {dx_i\over dl} 
  = v_i\Big(\sum\nolimits_j v_j v_j\Big)^{-1/2}\;,\quad
  v_i = - \sum\nolimits_j g^{-1}_{ij}{\partial\tau\over\partial x_j}\;.
\end{equation}
Here $g^{-1}_{ij}$ is the inverse matrix of the metric tensor $g_{ij}$
and $l$ is now the arc length with respect to
the non-Cartesian coordinate system ($dl^2\!=\!\sum_i dx_i dx_i$).

When $\mathcal{S}$ is discrete and the reaction is governed by a master 
equation with transition rates $k_{yx}$ (from state $x$ to state $y$),
the MFPT $\tau_x$ from state $x$ to the product 
$\mathcal{P}$ is again employed as a reaction coordinate.
But in order to determine reaction paths an analogue of metric is required,
as reaction paths for continuous systems are determined via gradients 
which involve metric.
The most obvious choice for an analogue of metric is
the transition rates $k_{yx}$ themselves,
and we suggest the scheme that a reaction path going through state $x$ 
chooses the next state $y$ such that $k_{yx}(\tau_x\!-\!\tau_y)$ is maximized.
For the transition step from $x$ to $y$, the transition time $1/k_{yx}$
may be interpreted as a cost, and the decrease $\tau_x\!-\!\tau_y$ in the MFPT
as a gain.
The scheme then amounts to maximizing the ratio between these two times,
namely the gain-cost ratio, for each step.

According to the above scheme, a reaction path is constructed starting from
each state in the reactant set $\mathcal{R}$.
In general, multiple reaction paths are obtained unless 
the reactant set is narrowed down to a single state.
Some reaction paths may overlap if they go through a common state.

\paragraph{\textbf{Examples.}}

To demonstrate the above scheme and its outcomes,
we consider first a single-particle Brownian motion on a two-dimensional 
potential surface described with a Cartesian coordinate system ($x_1,x_2$).
We take the three-hole potential 
\begin{eqnarray}
  U(x_1,x_2)&=&-3e^{-x_1^2-(x_2-5/3)^2}
  +3e^{-x_1^2-(x_2-1/3)^2}\nonumber\\
  &&-5e^{-(x_1-1)^2-x_2^2}-5e^{-(x_1+1)^2-x_2^2}
\end{eqnarray}
which was also studied by others regarding the temperature dependence of 
reaction paths~\cite{Huo1997,Elber2000}.
\begin{figure}
  \includegraphics{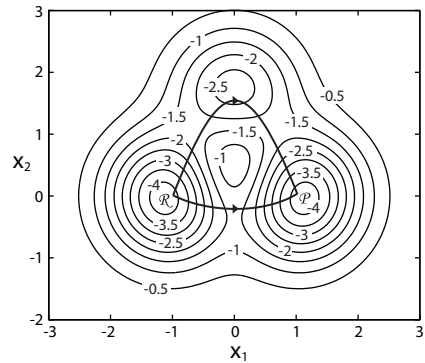}
\caption{\label{fig/pot} A contour plot of the three-hole potential, 
with two candidates for reaction path.}
\end{figure}
The potential features two deep holes and one shallow hole
(Fig.~\ref{fig/pot}).
The two deep holes are considered as the reactant and the product.
Roughly two possible pathways can be seen;
the upper path is longer than the lower one but has lower energy barriers.
It is therefore expected that the upper path will be taken at low temperature 
and that the lower path will be taken at high temperature.

The Brownian motion can be described in terms of the probability distribution
$p(\mathbf{x},t)$ and the probability current $\mathbf{J}(\mathbf{x},t)$.
In the strong friction regime, they satisfy the Smoluchowski 
equation~\cite{Gardiner1985}
\begin{eqnarray}
  &&{\partial\over\partial t}p(\mathbf{x},t)
  = -\nabla\cdot\mathbf{J}(\mathbf{x},t) \nonumber\\ 
  &&\mathbf{J}(\mathbf{x},t) 
  = -(\beta\gamma)^{-1}e^{-\beta U(\mathbf{x})}\nabla
  [e^{\beta U(\mathbf{x})}p(\mathbf{x},t)]\;,
\end{eqnarray}
where $\gamma$ is the friction coefficient.
The MFPT $\tau(\mathbf{x})$ 
is then obtained by solving the inhomogeneous partial differential equation
\begin{equation}
\label{tau}
  \nabla\cdot[e^{-\beta U(\mathbf{x})}\nabla\tau(\mathbf{x})]
  = -\beta\gamma e^{-\beta U(\mathbf{x})}\;
\end{equation}
with appropriate boundary conditions~\cite{MFPT}.
We take the region $(-4\!\le\!x_1\!\le\!4,-3\!\le\!x_2\!\le\!4)$ as the whole
set $\mathcal{S}$, and assume that its boundary is reflecting, namely
the probability current $\mathbf{J}$ is tangential to the
boundary.
For the reactant $\mathcal{R}$ we take the single point $(-1,0)$,
and for the product $\mathcal{P}$ we take the circular region of radius
0.1 centered at $(1,0)$.
Because a reaction event ends when the particle reaches the product region
$\mathcal{P}$, the boundary of $\mathcal{P}$ is absorbing, with
the probability distribution $p$ vanishing at the boundary.
Boundary conditions for $p$ and $\mathbf{J}$ lead to 
corresponding boundary conditions for $\tau$:
$\tau$ vanishes at absorbing boundaries and $\nabla\tau$ is tangential 
to reflecting boundaries~\cite{MFPT}.

\begin{figure}
  \includegraphics{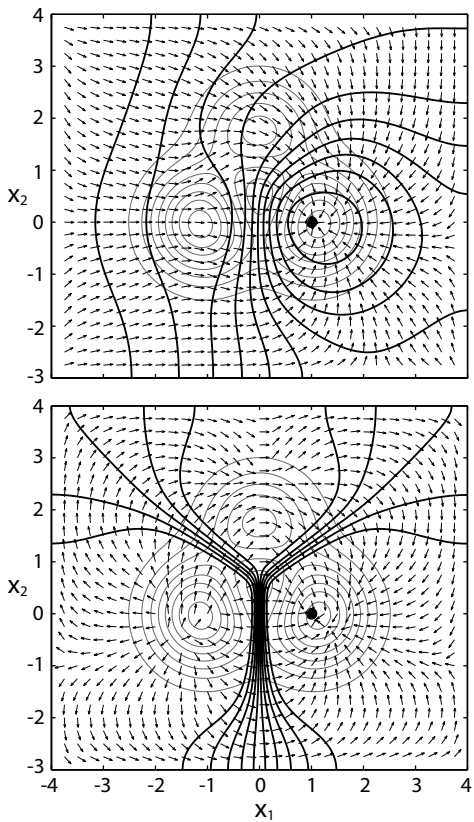}
\caption{\label{fig/arrow} Reaction paths of the Brownian motion
on the three-hole potential.
\textit{Top}, $\beta\!=\!1$; \textit{bottom}, $\beta\!=\!8$.
The directions of $-\nabla\tau$ at selected grid points 
are plotted as arrows.
Contours of the reaction coordinate $\tau$ (thick lines) 
and the potential (thin lines) are shown.
The product region is indicated by the filled circle.}
\end{figure}

The solutions of Eq.~(\ref{tau}), 
numerically obtained with MATLAB~\cite{Matlab}, 
for two different temperatures are shown in Fig.~\ref{fig/arrow}.
The differences between the two temperatures are dramatic.
At the high temperature ($\beta\!=\!1$) the arrows denoting the directions of
$-\nabla\tau$ flow more or less directly towards the product. 
At the low temperature ($\beta\!=\!8$), on the other hand, 
the flow is significantly distorted so that energy barriers are avoided,
with a singular point produced around $(-1.5,-0.5)$.
Also, the reaction coordinate $\tau$ drops rapidly when barriers are crossed, 
as indicated by the contours of $\tau$ packed around saddle points.

\begin{figure}
  \includegraphics{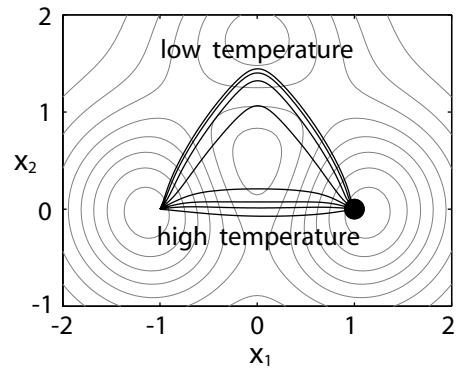}
\caption{\label{fig/paths} Temperature dependence of reaction paths 
of the Brownian motion on the three-hole potential.
Shown are eight reaction paths for eight different temperatures, 
from bottom to top, $\beta\!=\!1,2,3,4,5,6,7,8$.
The reactant is the point $(-1,0)$, and the product
is the region indicated by the filled circle.}
\end{figure}

Fig.~\ref{fig/paths} shows the reaction paths found at various temperatures.
As was expected, lower paths are taken at high temperature 
and upper paths are taken at low temperature.
At intermediate temperature, such as $\beta\!=\!4$, reaction paths lie 
in between, indicating that the upper and lower paths are equally favorable.

In order to demonstrate the scheme for reaction paths of discrete systems, 
we consider light-harvesting 
complexes~\cite{Blankenship2002,LH,Damjanovic2002A},
choosing photosystem I of cyanobacterium \textit{Synechococcus elongatus} 
as an example.
Photosystem I is a protein-pigment complex, embedded in the cell membrane, 
that contains 96 chlorophylls.
The aggregate of these chlorophylls is responsible for the first step 
in photosynthesis, namely absorption of light 
and subsequent migration of the resulting electronic excitation 
towards the special pair of chlorophylls, called P700 and located at the
geometrical center of the aggregate,
where the next step in photosynthesis, the charge separation, occurs.
This excitation migration can be considered a reaction.
Assuming that no more than a single chlorophyll is simultaneously excited
in the system, the states are specified by the excited chlorophylls.
The reactant is specified by the chlorophyll that initially absorbed a photon;
the product is specified the P700 pair of chlorophylls.
Reaction paths of this system provide
representative and minimal pathways of the excitation migration 
from the initially excited chlorophyll to the P700 pair.

\begin{figure} \includegraphics{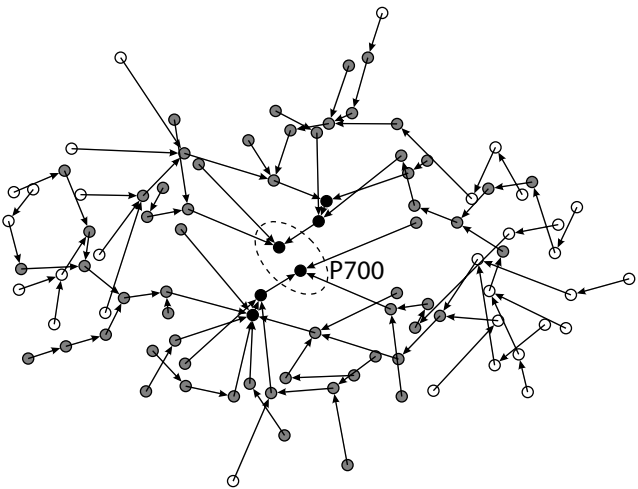}
\caption{\label{fig/ps1arrow}Reaction paths of the excitation migration
in photosystem I\@.
The positions of the chlorophylls, projected onto the membrane
plane, are denoted by circles colored according to the MFPT: 
black, short MFPT; gray, intermediate MFPT; white, long MFPT.}
\end{figure}

Since we are interested in first passage of excitation to P700, 
it is convenient to consider a subsystem of 94 chlorophylls,
excluding the P700 pair of chlorophylls.
The migration of excitation in this subsystem can be described 
by the master equation
\begin{equation}
\label{master}
  {d\over dt}p_x(t) = \sum\nolimits_yK_{xy}\,p_y(t)\;,
\end{equation}
where $p_x(t)$ is the probability that the excitation resides 
at chlorophyll $x$
at time $t$ and $K_{xy}$ is a $94\!\times\!94$ transition 
matrix~\cite{Ritz2001,Park2002A}.
We build the transition matrix $K_{xy}$ by using the inter-chlorophyll 
excitation transfer rates that were calculated in Ref.~\cite{Sener2002} 
based on the theory developed in~\cite{Damjanovic1999} and the
recently obtained structure of photosystem I~\cite{Jordan2001}.
The MFPT $\tau_x$ from chlorophyll $x$ to the P700 pair 
is then~\cite{Park2002A}
\begin{equation}
  \tau_x = -\sum\nolimits_y \phi_y K^{-1}_{yx}\Big/ \phi_x\;, \quad
  \phi_x = -\sum\nolimits_y \xi_y K^{-1}_{yx}\;.
\end{equation}
Here $K^{-1}_{xy}$ is the inverse matrix of $K_{xy}$ and $\xi_x$ is the 
excitation transfer rate from chlorophyll $x$ to P700.
The reaction paths constructed from the MFPT 
are shown in Fig.~\ref{fig/ps1arrow}.
A detailed discussion of this system is reported in~\cite{Park2002A}.

\paragraph{\textbf{Conclusion.}}

We have presented a new method to construct reaction paths based on the
MFPT gradient that incorporates all reaction events,
and illustrated how it captures important aspects of reactions, most notably
temperature effects.
Unlike previous attempts,
the present method provides paths starting from all states of the system.
We believe that the MFPT is the most natural choice for reaction coordinate, 
and expect that our approach will be used in many future studies of reactions.
The generality of MFPT permits applications of the present method 
to a wide range of phenomena.
The present method is particularly suitable for large reaction networks
that are completely characterized by the method 
through a pathway graph that connect each state to the product.

\begin{acknowledgments}
We thank Paul Grayson and Melih K.\ \c{S}ener for useful discussions.
This work has been supported by National Institutes of Health 
grant PHS 5 P41 RR05969.
\end{acknowledgments}

\end{document}